\documentstyle[12pt]{article}
\begin{document}

\title{The simplest Regge calculus model in the canonical
 form}
\author{V.M.Khatsymovsky \\
 {\em Budker Institute of Nuclear Physics} \\ {\em
 Novosibirsk,
 630090,
 Russia} \\ {\em E-mail address: khatsym@inp.nsk.su}}
\date{}
\maketitle
\begin{abstract}
Dynamics of a Regge three-dimensional (3D) manifold in a
continuous time is considered. The manifold is closed
consisting of the two tetrahedrons with identified
corresponding vertices. The action of the model is that
obtained via limiting procedure from the general relativity
(GR) action for the completely discrete 4D Regge calculus.
It closely resembles the continuous general relativity
action in the Hilbert-Palatini (HP) form but possesses
finite number of the degrees of freedom. The canonical
structure of the theory is described. Central point is
appearance of the new relations with time derivatives not
following from the Lagrangian but serving to ensure
completely discrete 4D Regge calculus origin of the system.
In particular, taking these into account turns out to be
necessary to obtain the true number of the degrees of
freedom being the number of linklengths of the 3D Regge
manifold at a given moment of time.
\end{abstract}
\newpage
Regge calculus recently remains
probably the most promising tool in the attempts to
quantise gravity \cite{Wil, Imm, Amb}. It's usual
application is that in the form of simplicial
minisuperspace \cite{Har} used in finding a wavefunction
of the universe by summation over histories in the Feynman
path integral \cite{Sil}. The advantage of the
minisuperspace is that it possesses a finite number of the
degrees of freedom. This makes path integral well-defined
as usual multiple integral.

In the simplicial minisuperspace framework the area
variables on the triangles seem to be more natural than
the linklengths \cite{Bar}; at the same time neighbouring
area bivectors are not independent. One can view the
long-distance effects such as gravity waves as a result of
correlations between the neighbouring areas for if these
were independent, the theory would be locally trivial by
Einstein eqs. in the Regge calculus framework. It was shown
in \cite{Roc} how Regge calculus reproduces gravity waves
of GR in the long wavelength limit for the periodic Regge
manifold of a specific form.

Now introduce the model consisting of the two evolving in
the continuous time tetrahedrons with common face $ABC$
and identified fourth vertices $D$ and $D^{\prime}$. This
is an abstract situation when the long distance effects
(gravity waves) are absent simply because the "long
distances" themselves are absent. Thereby we concentrate
on the pure local effects steming from the neighbourhood
of the only few triangles.

In the given paper we formulate this model in the canonical
form. It turns out that using Regge calculus in the area
bivector -- connection variables \cite{Kha1} leads to the
form of the action which notationally has almost complete
correspondence with the HP form of the
continuum GR action. The advantage of the connection
variables is that these are independent ones, at the same
time their being excluded from the action with the help of
eqs. of motion gives exactly Regge action in terms of
linklengths. The considered action is the limiting form of
the full 4D discrete Regge calculus action obtained by
passing to the continuous time. The continuous time
formalism provides the strict framework for the canonical
formalism \cite{Kha2} and subsequent canonical (Dirac)
quantisation.

Central point of the paper is appearance of the additional
constraints not present in the original Lagrangian and
simultaneous extension of the phase space necessary to
obtain the true number of the degrees of freedom. These
constraints are continuous time limit form of the relations
showing that definite bivectors correspond to triangles
which form 4D Regge manifold. It was shown in \cite{Kha2}
that the constraints of such type follow from the
Lagrangian in the case when the number of the triangles
$N_2$ is not smaller than the number of links $N_1$ in the
3D leaf and do not require to be set additionally in such
case. However, in our case $N_2$ = 4, $N_1$ = 6, and the
situation is different.

Let us first briefly repeat the formulas of \cite{Kha2} of
interest. If the triangle is spanned by two links with
vectors $l^a_1$, $l^a_2$ (in the local frame of one of the
tetrahedrons sharing this triangle) let us define the
bivector of the triangle via
\begin{equation}                                         
\pi_{ab}=\pm\varepsilon_{abcd}l^c_1l^d_2
\end{equation}
where $a$, $b$, $c$,... = 0, 1, 2, 3 are local indices,
$\varepsilon_{abcd}$ is completely antisymmetric tensor,
$\varepsilon_{0123}$ = +1. Let us denote the vertices of
the both tetrahedrons by $\mu$, $\nu$, $\lambda$,... = 0,
1, 2, 3 and, by the same symbols, the opposite triangles.
A sign in the definition of the bivector $\pi^\mu$ of the
triangle $\mu$ is chosen so that $\pi^0$ + $\pi^1$ +
$\pi^2$ + $\pi^3$ = 0. It is convenient to write
$v_\mu n_{\mu\nu}dt$ for the infinitesimal bivector of the
timelike triangle formed by the link $(\mu\nu)$ which
connects the vertices $\mu$ and $\nu$ and by the
infinitesimal timelike link $(\mu\mu^\prime)$ connecting
$\mu$ with it's image $\mu^\prime$ in the next time leaf
(taken at the moment $t+dt$ of the world time).
The $v_\mu$ defines the scale of (local)
time at the point $\mu$. In \cite{Kha2} the $\pi$ and
$n$ bivectors were shown in general case to be subject to
a system of bilinear constraints guaranteing their being
bivectors of some Regge manifold. In our simple case we can
write out explicit formula for $n_{\mu\nu}$ in terms of
$\pi^\mu$ and choose the time scale parameter $v_\mu$ so
that
\begin{equation}                                         
\label{n-mu-nu}
n_{\mu\nu}=\sum_{\lambda ,\rho}{
\varepsilon_{\mu\nu\lambda\rho}\left (u^\lambda_\mu\pi^\rho
+{1\over 2}[\pi^\lambda ,\pi^\rho]\right )}
\end{equation}
where $u^\lambda_\mu$ are parameters (independent
variables). The timelike triangle $(\mu\mu^\prime\nu)$ is
shared (in the 4D manifold) by the two tetrahedrons
$(\mu\mu^\prime\nu\lambda)$ and $(\mu\mu^\prime\nu\rho)$
where $\lambda$, $\rho$ are the two rest vertices other
than $\mu$, $\nu$. The two connection SO(3,1)
matrices (SO(4) in the Euclidean case) live on
these tetrahedrons and define the curvature matrix on the
triangle $(\mu\mu^\prime\nu)$ --- $R_{\mu\nu}$. The
connection on $(\mu\mu^\prime\nu\lambda)$ will be denoted
as $\Omega_{\mu\rho}$ with ordered pair of indices
$\mu\rho$ denoting the tetrahedron spanned by the link
$(\mu\mu^\prime)$ and triangle $\rho$ =
$(\mu\nu\lambda)$. Then
\begin{equation}                                         
R_{\mu\nu}={1\over 2}\sum_{\lambda ,\rho\neq\mu ,\nu}{
\left (\bar{\Omega}_{\mu\rho}
\Omega_{\mu\lambda}
\right )^{\varepsilon_{\mu\nu\lambda\rho}}}
\end{equation}
where overlining means Hermitean conjugation (in fact,
this reduces to only one term, e.g.
$R_{01}$ $=$ $\bar{\Omega}_{03}\Omega_{02}$). The
$\Omega_{\mu\nu}$ were shown \cite{Kha2} to be
parameterised as follows:
\begin{equation}                                         
\label{omega}
\Omega_{\mu\nu}=\Omega_\nu \exp{(\varphi^\nu_\mu\pi^\nu
+\,^*\!\varphi^\nu_\mu\,^*\!\pi^\nu)}
\end{equation}
where
\begin{equation}                                         
\,^*\!\pi_{ab}={1\over 2}\varepsilon_{abcd}\pi^{cd}
\end{equation}
and $\varphi^\nu_\mu$, $\,^*\!\varphi^\nu_\mu$ are
parameters (independent variables). The form (\ref{omega})
is necessary to ensure finiteness of the action: the
deficit angle which residues on spacelike triangle is
generally not infinitesimal, and (\ref{omega}) provides up
to $O(dt)$ cancellation of the contributions into action
from these triangles located inside the infinitesimal
prism with bases $(\mu\nu\lambda)$ and
$(\mu^\prime\nu^\prime\lambda^\prime)$; on the other hand,
namely the fact that exponential in (\ref{omega}) cannot
be absorbed into $\Omega_\nu$ for different $\mu$, i.e.
$\Omega_{\mu_1\nu}$ $\neq$ $\Omega_{\mu_2\nu}$ for
$\mu_1$ $\neq$ $\mu_2$, leads to independence between
$R_{\mu\nu}$ and $R_{\nu\mu}$: $R_{\mu\nu}$ $\neq$
$R^{\pm 1}_{\nu\mu}$, so that the planes of the triangles
$(\mu\mu^\prime\nu)$ and $(\mu\nu^\prime\nu)$ around
which $R_{\mu\nu}$ and $R_{\nu\mu}$ rotate do not coincide
and therefore the links $(\mu\mu^\prime)$ and
$(\nu\nu^\prime)$ (analogs of the lapse-shift vectors at
the different vertices) are independent of each other as
these should be. It is the remaining factor $\Omega_\nu$
which can be attributed to the triangle $\nu$ when we speak
of the geometry induced in the 3D leaf and we can speak of
it as that rotating from one tetrahedron to another one in
this leaf. Defining scalar product of bivectors $A^{ab}$
and $B^{ab}$ as
\begin{equation}                                         
A\circ B = {1\over 2}{\rm tr}A\bar{B}
\end{equation}
we can write out the Lagrangian:
\begin{equation}                                         
L = \sum_{\mu}{\pi^\mu\circ\bar{\Omega}_\mu
\dot{\Omega}_\mu} - \sum_{\mu}v_\mu H_\mu
- h\circ\sum_{\mu}{\pi^\mu}
- \tilde{h}\circ\sum_{\mu}{
\Omega_\mu\pi^\mu\bar{\Omega}_\mu}
\end{equation}
(dot means time derivative) where
\begin{eqnarray}                                         
H_\mu & = & \sum_{\nu}{|n_{\mu\nu}\!|\arcsin{
\frac{n_{\mu\nu}\circ R_{\mu\nu}}{|n_{\mu\nu}\!|}}
+ \varphi^\nu_\mu\pi^\nu\circ\Delta_\mu\pi^\nu}\\
\Delta_\mu\pi^\nu & \stackrel{\rm def}{=} &              
\sum_{\lambda ,\rho}{\varepsilon_{\mu\nu\lambda\rho}
n_{\mu\lambda}}
\end{eqnarray}
The $\Delta_\mu\pi^\nu$ defines variation
$v_\mu\Delta_\mu\pi^\nu dt$ of the bivector $\pi^\nu$ due
to the shifting the vertex $\mu$ to $\mu^\prime$. By
varying $L$ in the antisymmetric $h$ and $\tilde{h}$
we get the Gauss law in each tetrahedron expressing
closeness of the tetrahedron surface.

Now consider the constraints not following from the
Lagrangian. Among these there are the following ones:
\begin{equation}                                        
\label{pi-pi}
\pi^\mu *\pi^\nu\stackrel{\rm def}{=}\pi^\mu\circ
(\,^*\!\pi^\nu) = 0
\end{equation}
ensuring $\pi^\mu$ being face bivectors of some
tetrahedron. The four such constraints with $\mu$ = $\nu$
are, in fact, I class ones: these appear added to the
Lagrangian at the following symmetry transformation:
\begin{eqnarray}                                        
\Omega_\mu &\rightarrow &\Omega_\mu\exp{(\zeta^\mu\,^*
\!\pi^\mu)},\\
\,^*\!\varphi^\mu_\nu &\rightarrow &                    
\,^*\!\varphi^\mu_\nu - \zeta^\mu,
\end{eqnarray}
$\zeta^\mu$ being parameters. Note that the other I class
constraints, the Gauss law, serve to maintain invariance
at the usual local frame rotations:
\begin{eqnarray}                                        
\pi^\mu &\rightarrow & U\pi^\mu\bar{U},\\
\Omega_\mu &\rightarrow & O\Omega_\mu\bar{U},\\         
h &\rightarrow & h - \bar{U}\dot{U},\\                  
\tilde{h} &\rightarrow &\tilde{h} + \bar{O}\dot{O},     
\end{eqnarray}
$O, U\in$ SO(3,1) (SO(4) in the Euclidean case). These
constraints being I class can be displayed also by their
commutativity with other constraints (including the
Hamiltonian) w.r.t. the Poisson brackets (PB) defined for
any function $f$ of $\pi$, $\Omega$ so that
\begin{equation}                                        
\dot{f} = \{ H,f\}.
\end{equation}
This leads to the following definition:
\begin{equation}                                        
\{ f,g\} = \pi\circ [f_\pi ,g_\pi ]
+ f_\pi\circ\bar{\Omega}g_\Omega
- g_\pi\circ\bar{\Omega}f_\Omega.
\end{equation}
Here indices $\pi$, $\Omega$ mean corresponding derivative;
the derivatives over $\Omega$ times $\bar{\Omega}$ and over
$\pi$ are assumed to be symmetrised; the summation over the
pairs ($\pi^\mu$, $\Omega_\mu$) is implied.

The two independent constraints among (\ref{pi-pi}) with
$\mu$ $\neq$ $\nu$ are not I class. As those we can choose,
e.g., $\pi^1*\pi^2$ and $\pi^2*\pi^3$ (such as
$\pi^0*\pi^1$ and $\pi^2*\pi^3$ are not independent modulo
I class costraints).

Thus far we have 16 first class constraints (including the
12 Gauss law components) and 6 second class constraints
(including the four $H_\nu$). Without taking the
constraints into account the phase space of $\pi$, $\Omega$
would correspond to 24 degrees of freedom; taking the above
constraints into account diminishes this number to 24 - 16
- ${1\over 2}$6 = 5. It does not coincide with the number
of leaf linklengths which is 6. (Note that the nondynamical
variables $u^\nu_\mu$, $\varphi^\nu_\mu$,
$\,^*\!\varphi^\nu_\mu$ enter $L$ nonlinearly and are given
as implicit functions of $\pi$, $\Omega$ by eqs. of motion
for them,
\begin{equation}                                        
\frac{\partial{\cal H}_\mu}{\partial (u^\nu_\mu ,
\varphi^\nu_\mu , \,^*\!\varphi^\nu_\mu )} = 0
\end{equation}
).

This disagreement takes place because we have not
completely specified our continuous time system as limiting
case of the completely discrete 4D Regge manifold. Indeed,
whereas the constraints (\ref{pi-pi}) define the leaf
tetrahedron at the time $t$, and expression (\ref{n-mu-nu})
for $n_{\mu\nu}$ guarantees that the $n_{\mu\nu}$ are the
bivectors of rigid 4D structure filling in the spacetime
between the two successive leaves at $t$ and $t$ + $dt$,
there is also need in the relations ensuring possibility
of glueing together the two successive such structures: one
between $t$ and $t$ + $dt$ leaves and another one, say,
between $t$ + $dt$ and $t$ + $2dt$ leaves. As such
relations, one can set continuity of scalar $\pi^\mu$
bilinears on the junction leaf at $t$ + $dt$: on the one
hand, to form the boundary of the 4D submanifold between
$t$ and $t$ + $dt$, such bivectors should be
$$\pi^\mu (t) + \sum_{\nu}{v_\nu\Delta_\nu\pi^\mu}dt;$$ 
on the other hand, in the 4D submanifold between $t$ + $dt$
and $t$ + $2dt$ this is simply $\pi^\mu (t + dt)$. Thus,
\begin{equation}                                        
{d\over dt}(\pi^\mu\circ\pi^\nu)
= \Delta (\pi^\mu\circ\pi^\nu) (\Delta
\stackrel{\rm def}{=}\sum_{\nu}{v_\nu\Delta_\nu}).
\end{equation}
At $\mu$ = $\nu$ such relation follows from the eqs. of
motion, namely, gets added to the Lagrangian upon the
following variation of variables:
\begin{eqnarray}                                        
\Omega_\mu &\rightarrow &
\Omega_\mu\exp{(\xi^\mu\pi^\mu)},\\
\varphi^\mu_\nu &\rightarrow &                          
\varphi^\mu_\nu - \xi^\mu.
\end{eqnarray}
As the rest independent (modulo Gauss law) relations may
be chosen, e.g., those for $d(\pi^1\circ\pi^2)/dt$ and
$d(\pi^2\circ\pi^3)/dt$.

In principle, one can exclude the derivatives
$\dot{\pi^\mu}$ with the help of eqs. of motion thus
converting these relations into constraints on $\pi$,
$\Omega$ in the usual sense. But this introduces a
nonequivalence of time and space additional to the already
existing nonequivalence connected with continuity of time
and discreteness of space. Aiming at formulating the theory
maximally symmetrically we choose to add these relations to
the Lagrangian with the help of Lagrange multipliers,
$\psi_{12}$ and $\psi_{23}$, just as we (implicitly) do so
for other bilinear constraints on bivectors not containing
the time derivatives. The new Lagrangian up to the full
derivative reads:
\begin{equation}                                        
{\cal L} = L + \dot{\psi}_{12}\pi^1\circ\pi^2
+ \dot{\psi}_{23}\pi^2\circ\pi^3
+ \psi_{12}\Delta (\pi^1\circ\pi^2)
+ \psi_{23}\Delta (\pi^2\circ\pi^3).
\end{equation}
Thus we get the new dynamical variables. In the Hamiltonian
formalism the phase space is extended by the two new
canonical pairs ($\tilde{\psi}^{12}$, $\psi_{12}$) and
($\tilde{\psi}^{23}$, $\psi_{23}$), $\tilde{\psi}^{\mu\nu}$
being conjugate momenta subject to the two new constraints:
\begin{equation}                                        
\label{psi-pi-pi}
\tilde{\psi}^{12} - \pi^1\circ\pi^2 = 0,~~~
\tilde{\psi}^{23} - \pi^2\circ\pi^3 = 0.
\end{equation}
Besides, Hamiltonian constraints are modified:
\begin{equation}                                        
H_\mu \rightarrow {\cal H}_\mu = H_\mu - \psi_{12}
\Delta_\mu (\pi^1\circ\pi^2) - \psi_{23}
\Delta_\mu (\pi^2\circ\pi^3).
\end{equation}
The PB for the case of $\psi$, $\tilde{\psi}$ dependence
are modified in obvious way. The ${\cal H}_\mu$,
$\pi^1*\pi^2$, $\pi^2*\pi^3$ and new constraints
(\ref{psi-pi-pi}) generally form the second class system
since determinant of their PB is not identical zero. The
overall effect of extending phase space is therefore
enhancing the number of the degrees of freedom by 1, so
that this number is just 6, the number of leaf linklengths.

The two variables $\psi_{\mu\nu}$ can be excluded solving
two of the four Hamiltonian constraints ${\cal H}_\mu$.
This leaves us with only two
Hamiltonian constraints on the phase space of purely
($\pi$, $\Omega$) pairs.

The system obtained strongly remind the usual HP form of
the continuum GR (see, e.g., review \cite{Pel}). If (purely
formally, forgetting for a moment that the system is
strongly nonperturbative) we expand the action over small
$\omega_\mu$, $\varphi^\mu_\nu$, $\,^*\!\varphi^\mu_\nu$
(where $\Omega_\mu$ = $\exp{\omega_\mu}$), we reproduce
the kinetic term $\pi^\mu\circ\dot{\omega}_\mu$; analog of
the vector (coefficients at $u^\mu_\nu$),
Hamiltonian (coefficients at $v_\mu$) and Gauss law
constraint (coefficients at $h$, $\tilde{h}$); the
constraint ensuring the tetrad form of $\pi^\mu$
($\pi^\mu*\pi^\nu$ = 0). The difference is that, on the one
hand, differentiating the constraint $\pi^\mu*\pi^\nu$ now
does not yield the new constraint (rather defines some
Lagrange multipliers) and, on the other hand, the new
variables $\psi_{\mu\nu}$, $\tilde{\psi}^{\mu\nu}$ and
related new constraints arise. Besides that, analogs of the
vector and Hamiltonian constraints are not first class (do
not commute mutually and with other constraints).

\bigskip
This work was supported in part by the RFBR grant
No. 96-15-96317.

\end{document}